\def\preprint{1}            

\ifdefined\preprint
  \documentclass[preprint,review,12pt]{elsarticle}
\fi
\ifdefined\wordcount
  \documentclass[final,3p,times,twocolumn]{elsarticle}
\fi
\ifdefined\final
  \documentclass[final,3p,times,twocolumn]{elsarticle}
\fi

\usepackage{graphicx,stfloats}
\usepackage{color}
\usepackage{balance}

\usepackage{latexsym}
\usepackage{amssymb}
\usepackage{amsmath}
\usepackage{amsfonts}
\usepackage{xcolor}
\usepackage{mathrsfs}

\usepackage{hyperref}
\hypersetup{
  colorlinks   = true, 
  urlcolor     = blue, 
  linkcolor    = blue, 
  citecolor   = red 
}

\biboptions{sort&compress}

\journal{Proceedings of the Combustion Institute}

\begin{document}

\begin{frontmatter}

\title{Flame dynamics during intermittency and secondary bifurcation to longitudinal thermoacoustic instability in a swirl-stabilized annular combustor}

\author[1]{Amitesh Roy\corref{cor1}}
\cortext[cor1]{Corresponding author.}
\ead{amiteshroy94@yahoo.in}
\author[1]{Samarjeet Singh}
\author[1]{Asalatha Nair}
\author[2]{Swetaprovo Chaudhuri}
\author[1]{Sujith R I}
\address[1]{Department of Aerospace Engineering, IIT Madras, Tamil Nadu - 600 036, India}
\address[2]{Institute of Aerospace Studies, University of Toronto, Ontario - M3H 5T6, Canada}

%
%

\begin{abstract}
In this experimental study on a laboratory-scale turbulent annular combustor with sixteen swirl-stabilized burners, we study the flame-flame and flame-acoustic interactions during different dynamical states associated with the {\color{black}longitudinal mode} of the combustor. We simultaneously measure the acoustic pressure and CH* chemiluminescence emission of the flame using a high-speed camera. Upon increasing the equivalence ratio, the combustor undergoes the following sequence of transition: combustion noise (CN) to low amplitude {\color{black}longitudinal} thermoacoustic instability (TAI) through the state of intermittency (INT), and from low amplitude to high amplitude {\color{black}longitudinal} TAI through a secondary bifurcation. We report the first evidence of secondary bifurcation from low amplitude TAI to high amplitude TAI for a turbulent thermoacoustic system {\color{black}which allows us to test the flame response at two different amplitude of perturbation in a natural setting}. We find a significant difference in the dynamics of the flame interactions during the periodic part of intermittency and low and high amplitude TAI. Specifically, during the periodic part of intermittency, the phase difference between the local heat release rate (HRR) measured from various burners show significant phase slips in time. During low amplitude TAI, there are fewer phase slips among the HRR response of the burners, which result in a state of weak synchronization among the flames. During high amplitude TAI, we find that the flames are in perfect synchrony amongst themselves and with the pressure fluctuations. We then quantify the degree of temporal and spatial synchronization between different flames, and flames and pressure fluctuations using the Kuramoto order parameter and the phase-locking value. We show that synchronization theory can be conveniently used to characterize and quantify flame-acoustic interactions in an annular combustor.
\end{abstract}

\begin{keyword}
Annular combustor \sep Thermoacoustic instability \sep Flame-Flame Interaction \sep Secondary bifurcation \sep Synchronization

\end{keyword}

\end{frontmatter}

\ifdefined \wordcount
\clearpage
\fi


\section{Introduction}
\label{Introduction}
Gas turbine combustors typically utilize an annular arrangement of burners to facilitate continuous and spatially distributed combustion along the annulus. The coupling between the unsteady heat release rate (HRR) from the flames along the annulus with the acoustic pressure fluctuations can lead to self-excited longitudinal or transverse instability or a combination of the two. During unstable operation in an annular combustor, a large number of interactions take place concomitantly: Turbulent flow interacts with the premixed flames; flames interact with neighbouring flames; the flow and the flame interact with the acoustic field of the combustor \cite{o2015transverse,candel2014dynamics}.

The interaction of the neighbouring flames in an annular combustor results in complex three-dimensional flame dynamics. The structure of the interacting flame undergoes many changes depending upon the inter-flame distance and the flame holding characteristics. \citet{worth2012cinematographic} analyzed the effect of the separation distance between flames in an arrangement with two bluff-body stabilized flames. They concluded that lower inter-burner distances lead to large scale flame merging, resulting in an altered mean flame structure and their associated thermoacoustic response. In a follow-up study on a full annular burner, \citet{worth2013self} showed that the flame structure changed from helical to a large-scale merged flame structure when the inter-burner distance was decreased. Later, in a swirl-stabilized annular burner, \citet{bourgouin2013self} analyzed the modal dynamics associated with HRR perturbation during longitudinal and transverse instability. {\color{black}During longitudinal instability, they found that the flame dynamics showed some degree of desynchrony.} The effect of swirl on the interaction between neighbouring flames was analyzed recently in a three swirl-stabilized configuration by \citet{vishwanath2018experimental}. They found that the difference in swirl number between neighbouring flames can preferentially suppress or enable the formation of vortex breakdown bubbles. {\color{black}Finally, in a conceptual study, \citet{manoj2019synchronization} showed that four diffusion flames in ambient conditions could show complex dynamical states. Depending upon the distance between the flames, the flame behavior can switch between in-phase oscillations, anti-phase oscillations, state of amplitude death, or any combination of these. Thus, it follows that the flame-flame interaction plays a significant part in dictating the overall thermoacoustic response of annular combustors.}

Quite a few experimental \cite{worth2013modal, bourgouin2013self, bourgouin2015characterization} and theoretical \cite{noiray2013dynamic, ghirardo2013azimuthal} studies have contributed immensely towards our understanding of longitudinal and transverse instability in annular combustors. However, most of these studies individually assess the state of transverse or longitudinal thermoacoustic instability. Studies that capture the dynamical transition to thermoacoustic instability (TAI) in annular combustors through smooth variation of parameters remain few. {\color{black}One such notable study is that of \citet{prieur2017hysteresis}, where the authors mapped the various combustor dynamics on the parametric plane of equivalence ratio and bulk-flow velocity. They observed longitudinal and transverse instability. They identified that variation of equivalence ratio in the fuel-rich limit led to a hysteresis cycle with the combustor dynamics changing from chugging to spinning to standing transverse mode.}

In this study, we quantify the flame-flame and flame-acoustic interaction in a swirl-stabilized annular combustor consisting of sixteen burners during the transition from combustion noise (CN) to {\color{black}longitudinal} high amplitude TAI. {\color{black}There is a transition from CN to low amplitude TAI through the state of intermittency, and secondary bifurcation from low amplitude TAI to high amplitude TAI when equivalence ratio is increased from fuel-lean conditions. The presence of the state of low amplitude TAI and high amplitude TAI allows us to evaluate the nonlinear dependence of flame response to acoustic perturbation of low and high amplitude in a natural setting without resorting to external forcing. We compare and contrast the effect of these different amplitude oscillations on the global flame structure. We assess the local flame response by determining the normalized amplitude and phase of the heat release rate fluctuations of each burner during each of the above-mentioned dynamical states. Most importantly, we compare the response between a neighbouring pair of burners and evaluate the degree of mutual synchronization amongst them using the phase-locking value (PLV). We further determine the degree of synchrony between the heat release rate response of individual burners with the acoustic response of the combustor. Finally, we find the extent of spatial synchronization between all the burners during different dynamical states through the use of Kuramoto order parameter. We conclude that even for the relatively simple case of longitudinal instability where multiple flames are subjected to constant amplitude perturbations, the flame response remains non-trivial and unlike anything that has been reported till now.} 

\section{Experimental setup and measurements}
\label{2. Experimental setup and measurements}
\begin{figure}[t!]
\centering
\includegraphics[width=0.75\textwidth]{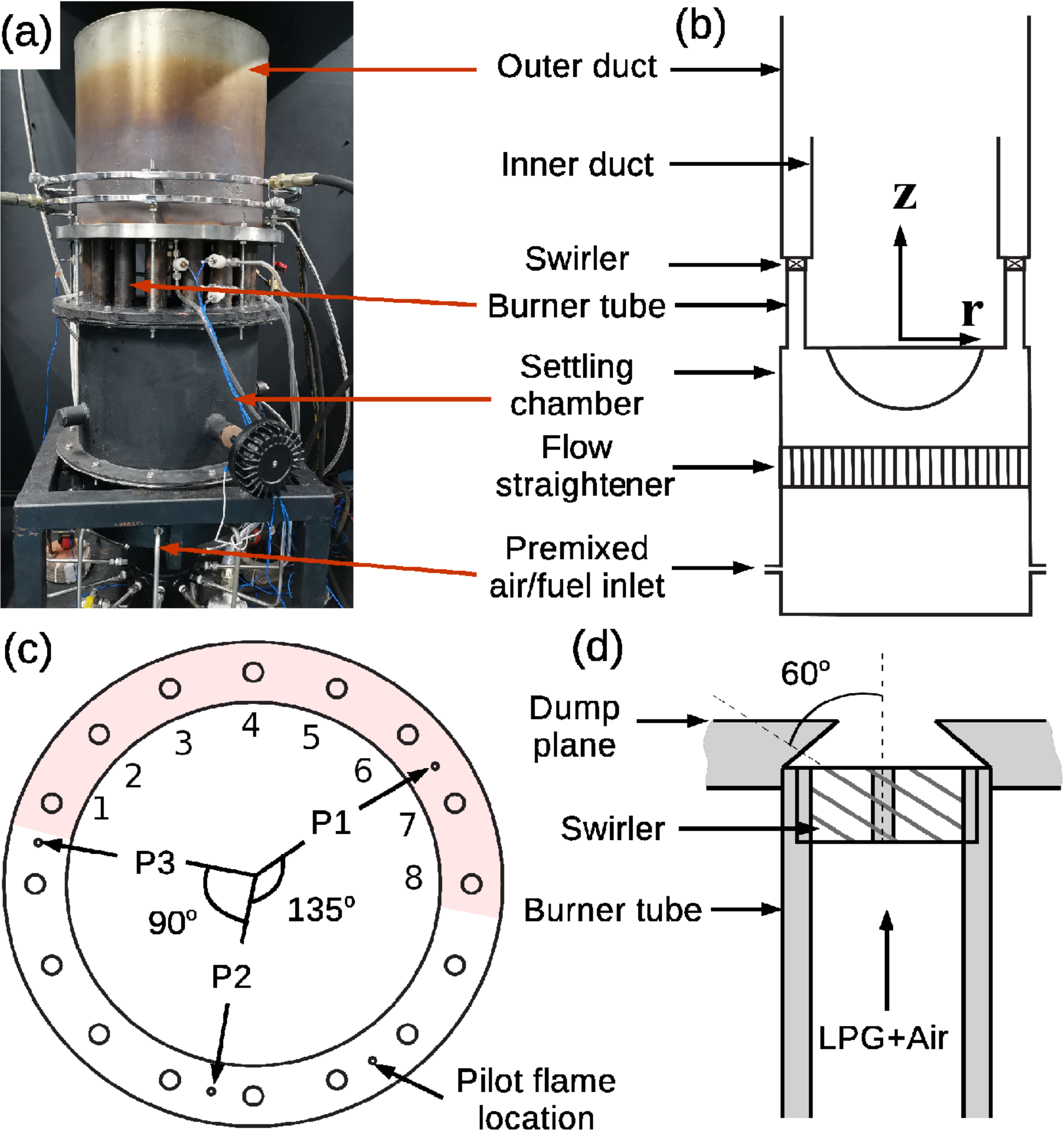}
\caption{\label{Fig1_Experimental_setup} (a) Front view of the annular burner. Schematic of the (b) combustor cross-section and (c) the top view. The half-plane which was imaged have been shaded and the burners serialized. (d) Schematic showing the burner with the swirler followed by a converging section.}
\end{figure}

The premixed annular combustor is shown in Fig. 1. The design of the annular combustor is inspired by the designs of \citet{worth2013modal} and \citet{bourgouin2013self}. The inner and outer diameter of the annulus is 300 mm and 400 mm, respectively. The lengths of the inner and outer ducts are 200 mm and 400 mm, respectively. There are sixteen burner tubes mounted on the annulus. The inner diameter and length of the burner tubes are 30 mm and 150 mm, respectively. Sixteen axial swirlers are mounted on each of the burners to impart solid-body counter-clockwise rotation downstream of the swirler. Each swirler consists of six guide vanes mounted on a central shaft of diameter 15 mm and inclined $\beta=60^o$ with respect to the injector axis. The geometric swirl number is $S=2/3\tan \beta=1.15$ \cite{candel2014dynamics}. A converging section with exit diameter $d=15$ mm connects the swirler to the annulus. The height of the converging section is 18 mm and has a contraction area ratio of 2 (Fig. \ref{Fig1_Experimental_setup}d). {\color{black} The separation distance between burners is $S=4.58d$.}

The sixteen burner tubes are connected to a settling chamber of diameter 400 mm and length 440 mm. {\color{black}Technically premixed air and} liquefied petroleum gas (LPG, 40\% propane and 60\% butane by volume) enters through the bottom of the settling chamber through 12 inlet ports, each having an internal diameter of 9.5 mm mounted perpendicular to the axis of the combustion chamber. The settling chamber contains flow straightener to arrest transverse velocity fluctuations. A flow divider is present to distribute the flow uniformly to each of the burners. 

Air and fuel flow rates are controlled using Alicat scientific mass flow controllers (MCR 2000SLPM for air and MCR 100SLPM for fuel). The equivalence ratio is varied by keeping the air flow rate constant and varying the fuel flow rate. Thus, $\phi$ is varied in the range of $0.3-0.6$ for nominal flow velocity $\upsilon_z\approx 8.5$ m/s and $Re_d\approx 8600$, respectively. {\color{black} The variation in the axial flow velocity, measured using a pitot tube at the centre of each burner at a downstream distance $z=10$ mm, is $8.60\pm 0.22$ m/s for $\upsilon_z\approx8.5$ m/s. Similarly, the axial velocity between consecutive burners at $z=10$ mm is $-2.33\pm0.17$ m/s. The negative flow velocity between burners indicates the presence of recirculation zones. The relatively low variation in the nominal flow velocity at the centre and between burners indicate spatially uniform flow across the annulus (see supplemental Fig. S1).} The maximum uncertainty in the values of $\phi$ is $\pm1.6\%$ and for $\upsilon_z$ and $Re$ is $\pm0.8\%$. The premixed flame is ignited using a non-premixed LPG pilot flame anchored between two injectors (Fig. \ref{Fig1_Experimental_setup}c).

Simultaneous pressure measurements and imaging were performed to acquire the acoustic pressure fluctuations and intensity fluctuations caused by the swirling flames. The acoustic pressure fluctuations are recorded using four PCB103B02 piezoelectric transducers (sensitivity - 217.5 mV/kPa, uncertainty - $\pm 0.15$ Pa). Three transducers are mounted {\color{black}on semi-infinite waveguides (diameter 4 mm and length 3.2 mm) at a distance of 75 mm from the combustor backplane. The waveguides are open to atmosphere. The location of the three transducers is indicated as P1, P2, and P3 in Fig. 1c}. The pressure signals are acquired for 3 s at a sampling frequency of 10 kHz and digitized using a National Instruments 16-bit PCI 6343 card. {\color{black} All the pressure signals shown in the results section are acquired from pressure transducer P1. All the results in this study remain invariant of the choice of the pressure sensor as we study longitudinal instability which entails no phase difference between the signals acquired by the three pressure sensors.}

A high-speed CMOS camera (Phantom V 12.1) is used to acquire the images at a resolution of $1280\times800$ pixels corresponding to the half-plane of the annulus of size $400\times200$ mm at full exposure. Imaging is performed with the aid of an air-cooled mirror placed overhead of the combustor. A CH* bandpass filter (bandwidth of $435\pm10$ nm) was used to capture chemiluminescence images of the flames. The camera is outfitted with a Nikon AF Nikkor 70-210 mm $f$/4-$f$/5.6 camera lens. A total number of 22,253 and 5,563 images were acquired during intermittency and TAI at a sampling frequency of 2000 Hz, respectively. A pulse generated using Tektronix AFG1022 function generator is used to trigger the camera and the PCI card to acquire measurements simultaneously. 

\section{Results and discussions}
\label{3. Results and discussions}

\subsection{Bifurcation diagram}
\begin{figure}[t!]
\centering
\includegraphics[width=0.75\textwidth]{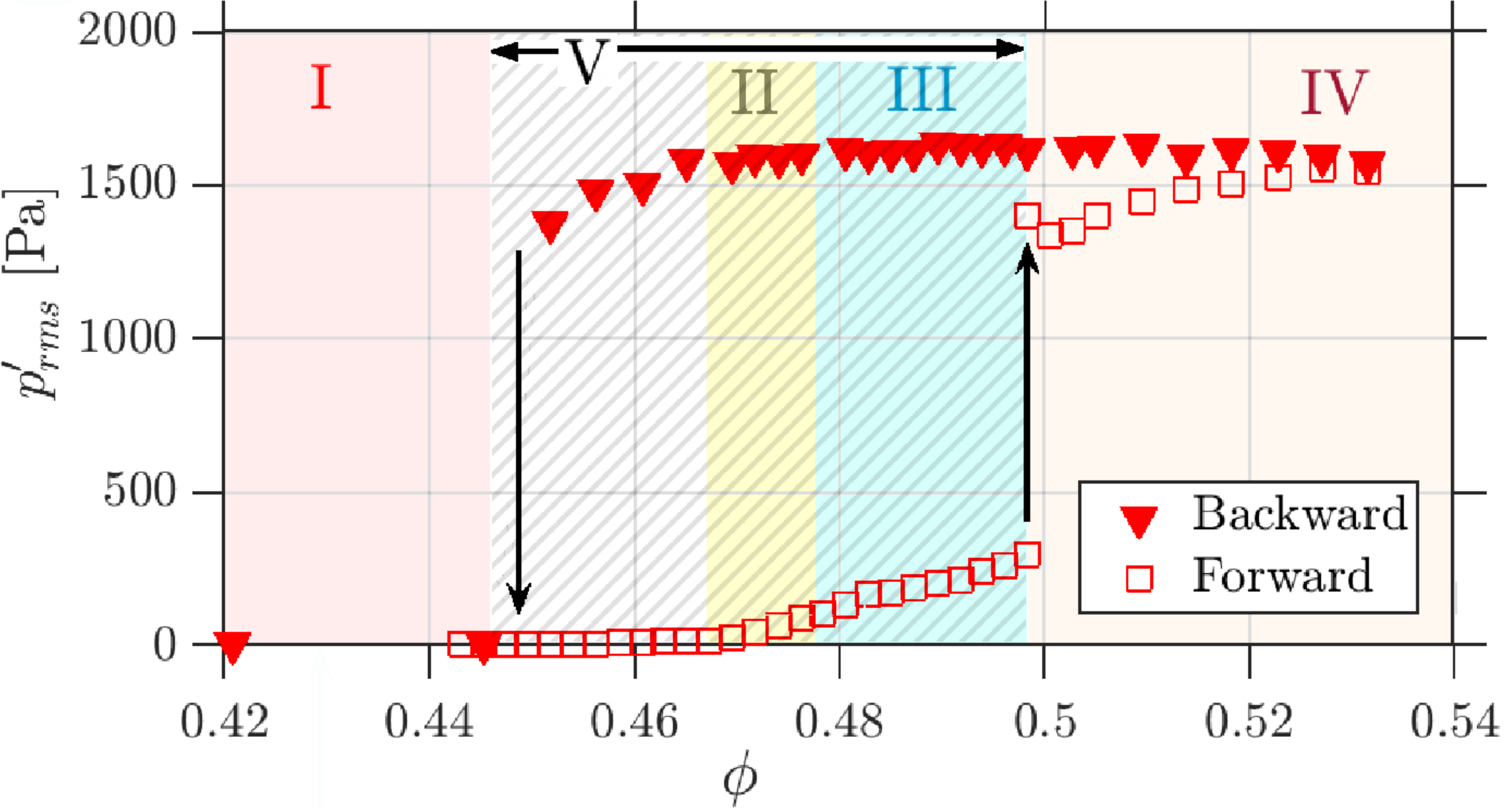}
\caption{\label{Fig2_prms_vs_phi} Observed transition to the state of TAI in the annular combustor when the equivalence ratio is increased (forward) and then decreased (backward). Region-I, II, III, and IV indicate the parametric regions where we observe the states of CN, INT, low amplitude TAI, high amplitude TAI, respectively. The hatched region (V) indicates the bistable region.}
\end{figure}

In Fig. \ref{Fig2_prms_vs_phi}, we plot the variation in the root-mean-square value of the acoustic pressure fluctuations ($p^\prime_{rms}$) as a function of the equivalence ratio $\phi$. For $\phi\lesssim 0.47$, the acoustic pressure fluctuations are aperiodic with a broadband amplitude spectrum and have a very low value of $p^\prime_{rms}$. This state of combustor operation is referred to as combustion noise (CN). Increasing $\phi$ leads to a state wherein low amplitude periodic oscillations are interspersed randomly amongst very low amplitude aperiodic oscillations. This state is referred to as intermittency (INT) \cite{nair2014intermittency}. 

\begin{figure*}[t!]
\centering
\includegraphics[width=\textwidth]{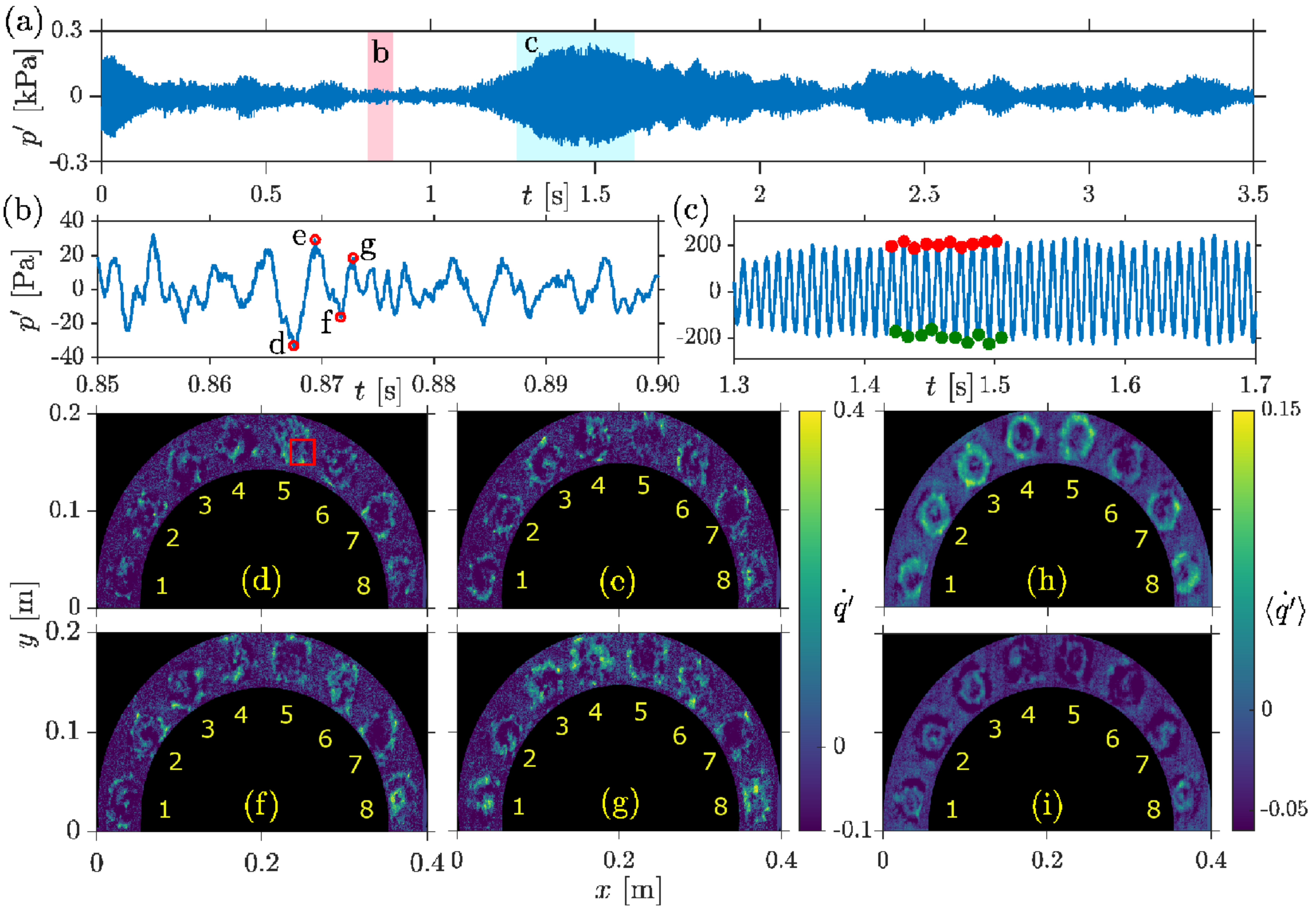}
\caption{\label{Fig3_image_int}(a) Time series of $p^\prime$ obtained during intermittency observed at $\phi=0.47$. (b) Aperiodic and (c) periodic epochs of intermittency. (d-g) Mean-subtracted instantaneous chemiluminescence images corresponding to the indicated points in the aperiodic region in (b). (h,i) Phase-averaged chemiluminescence image at the pressure maxima ($90^\circ$) and minima ($270^\circ$) measured from the points indicated in (c).}
\end{figure*}

When $\phi$ is increased past 0.48, there is a gradual increase in the sound level in the combustor. We observe periodic oscillations with a narrowband peak at $220\pm10$ Hz and amplitude levels of the order of $p^\prime_{rms}\sim 10^2$ Pa ($\approx135$ dB). We refer to this state as low amplitude TAI. At $\phi\approx0.5$ (Fig. \ref{Fig2_prms_vs_phi}), we observe an abrupt increase in the amplitude of pressure oscillations as the system dynamics transition from low amplitude TAI to high amplitude TAI of the order of $p^\prime_{rms}\sim 10^3$ Pa ($\approx 165$ dB). We also observe hysteresis when $\phi$ is decreased while the system is in the state of high amplitude TAI. Thus, the transition follows: CN (region-I) $\rightarrow$ INT (region-II) $\rightarrow$ low amplitude TAI (region-III) $\rightarrow$ high amplitude TAI (region-IV). {\color{black}Note that during high amplitude TAI, the sound intensity is nearly 30 dB larger than what is observed during low amplitude TAI. Such a secondary bifurcation have been predicted in nonlinear thermoacoustic system \cite{ananthkrishnan1998application}. To the best of our knowledge, this is the first experimental observation of secondary bifurcation in a turbulent thermoacoustic system. The secondary bifurcation occurs when sixth-order nonlinearities dominate the flame response. Such a modelling approach have been considered recently in \cite{singh2020intermittency}.}

{\color{black}The observation of secondary bifurcation is of great significance. Longitudinal instability stipulates that all the burners are subjected to pressure oscillations of similar amplitude. Thus, the flame response at each of the burners is expected to be nearly similar. We further know that flame has a nonlinear dependence on the amplitude of perturbation \cite{fleifil1996response}. Observation of limit cycle oscillation of two different amplitudes allows us to evaluate the nonlinear dependence of flame response at two different amplitudes in a natural setting without having to resort to external forcing. Next, we compare and contrast the flame response observed during these different dynamical states.}

\subsection{Flame dynamics during different dynamical states}
\subsubsection{Intermittency}
During CN, as the flames are subjected only to broadband turbulent velocity fluctuations, the HRR field largely remains incoherent, and the pressure fluctuations remain aperiodic and have not been shown here for brevity.

We focus on the flame dynamics observed during INT. The intermittent acoustic pressure oscillations observed when $\phi=0.47$ are shown in Fig. \ref{Fig3_image_int}a. In the enlarged portion in Figs. \ref{Fig3_image_int}b \& c, we can observe aperiodic and periodic pressure oscillations. Instantaneous mean-subtracted CH* images corresponding to the points in Fig. \ref{Fig3_image_int}b has been shown in Figs. \ref{Fig3_image_int}d-g. For the periodic part of intermittency, phase-averaged CH* images at maxima ($90^\circ$) and minima ($270^\circ$) determined from the red and green points in Fig. \ref{Fig3_image_int}b, have been shown in Figs. \ref{Fig3_image_int}h \& i. 

{\color{black}Fig. \ref{Fig3_image_int}d-g corresponds to local minima and maxima of aperiodic pressure oscillations. The intensity levels or the spatial distribution of HRR is incoherent across different burners, as is expected during aperiodic oscillations. Such an incoherent, desynchronized and non-uniform flame structure is also observed during CN.} In contrast, from the phase-averaged image taken at the pressure maxima during the periodic epoch of INT (Fig. \ref{Fig3_image_int}h), we can distinguish the swirling flame structure. For all the burners, we observe that the intensity is maximum along the periphery of the swirling flame. In comparison, the phase-averaged HRR field during the pressure minima shows negative values along the periphery of the flame, indicating flames annihilation events.

Now, we analyze the local HRR dynamics during the periodic part of intermittency. The local HRR is determined by summing over all the intensity value present in a rectangular region, as shown for the fifth flame in Fig. \ref{Fig3_image_int}d. The local region was taken instead of the entire burner to avoid phase cancellation effects from affecting the HRR time series. A similar region is chosen for all the burners and time series of the local HRR fluctuations is obtained. Since the local HRR signals contain phase noise, we bandpass the signal centred around the frequency of dominant oscillations ($f_n$) with a width of $\pm f_n/4$. Here, $f_n$ is the frequency of the limit cycle oscillations, which is approximately around $220\pm10$ Hz. 

{\color{black} We obtain the instantaneous phase of the HRR signal and the normalized time series based on the concept of analytic signals \cite{gabor1946theory}. We construct the analytic signal $\zeta(t)=\dot{q}^\prime_k(t)+i\mathscr{H}[\dot{q}^\prime_k]=A_k(t)\exp(i\theta_k t)$, where, $\dot{q}^\prime_k$ is the HRR signal for the $k^\text{th}$ burner, $A_k(t)$ is the instantaneous amplitude and $\theta_k$ is the instantaneous phase of the signal. The Hilbert transform is defined as:
\begin{equation}
\mathscr{H}[\dot{q}^\prime_k(t)]=\text{PV}\int_{-\infty}^\infty \dot{q}^\prime_k(\tau)/(t-\tau)d\tau,
\end{equation}
where PV indicates that the integral is evaluated at the Caucy principal value. The normalized HRR can then be determined as: $\dot{q}^\prime_k(t)/A_k(t)=\sin\theta_k(t)$ \cite{manoj2019synchronization}. Two oscillators ($\dot{q}^\prime_{i,j}$) are then said to be in phase synchronization if the phase difference between signals is constant i.e., $|\Delta\theta_{i,j}(t)|=|\theta_i-\theta_j|=$ constant.} 

\begin{figure}[t!]
\centering
\includegraphics[width=0.75\textwidth]{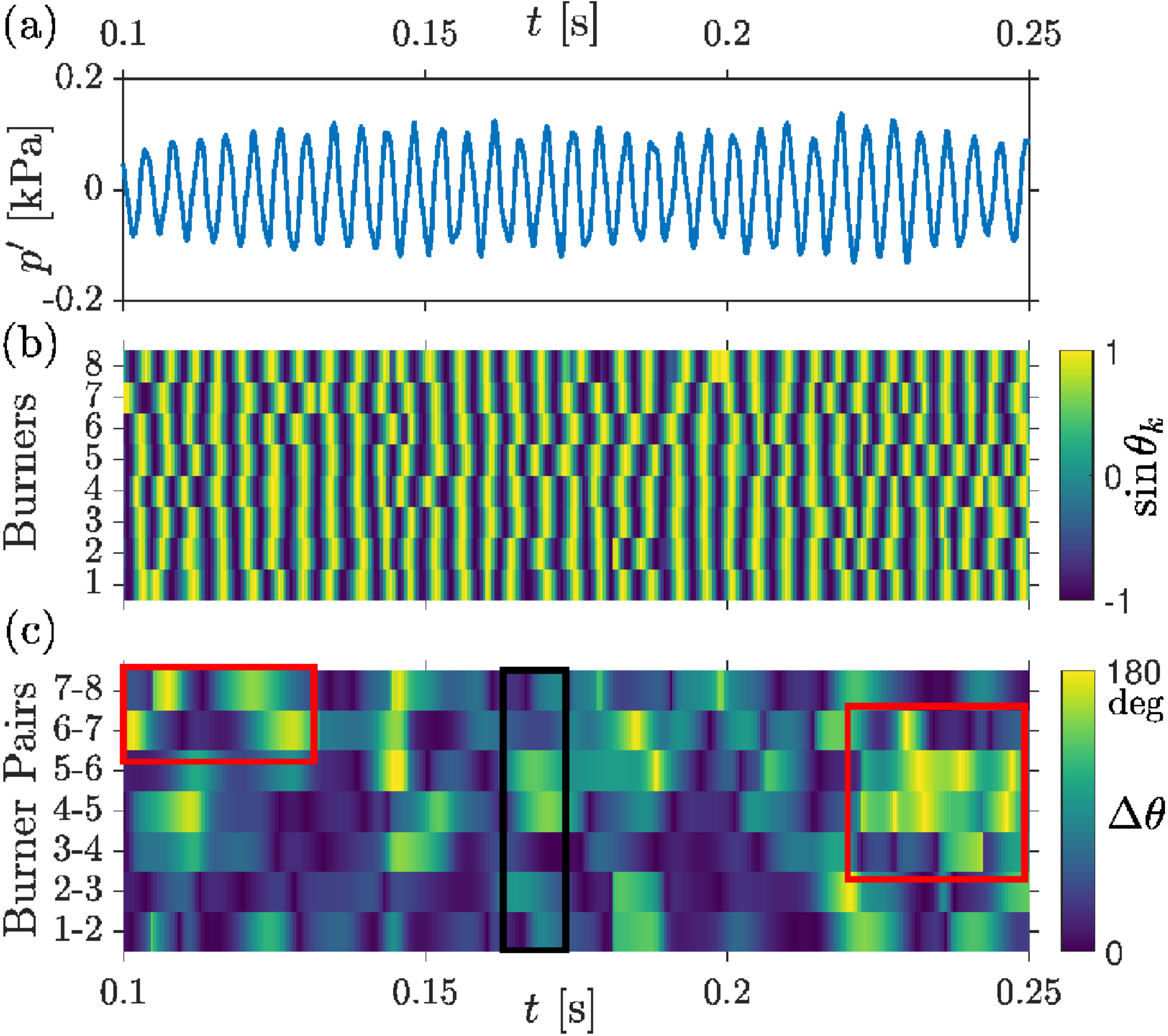}
\caption{\label{Fig4_P_vs_t_Amplitude_phase_diff_plot}(a) Periodic part of intermittency observed at $\phi=0.47$. Temporal variation of (b) the normalized amplitude ($\sin\theta$) of different burners and (c) the phase difference, $\Delta \theta$, between $\dot{q}^\prime$ between the burner pairs.}
\end{figure}

In Fig. \ref{Fig4_P_vs_t_Amplitude_phase_diff_plot}a, we show the periodic part of intermittency. In Fig. \ref{Fig4_P_vs_t_Amplitude_phase_diff_plot}b, we plot the temporal variation of the normalized amplitude of HRR oscillations ($\sin\theta_k(t)$) for all the burners. The phase difference between HRR oscillations of different pairs of burners is shown in Fig. \ref{Fig4_P_vs_t_Amplitude_phase_diff_plot}c. We observe a significant phase mismatch between the cycles of oscillations among different burners. The phase difference between neighbouring burners varies across the annulus. For instance, in the region indicated by the black rectangle, burner 1-2 are in-phase, while burner pair 4-5 is approximately $150^\circ$ out-of-phase. Such out-of-phase burner pairs are quite common as indicated by the red rectangle. {\color{black}Thus, even though the burners are frequency synchronized, they have significant phase desynchrony in time. In other words, the flames are in a state of partial (intermittent phase) synchronization with each other.}

\subsubsection{Low amplitude thermoacoustic instability}
\begin{figure}[t!]
\centering
\includegraphics[width=0.75\textwidth]{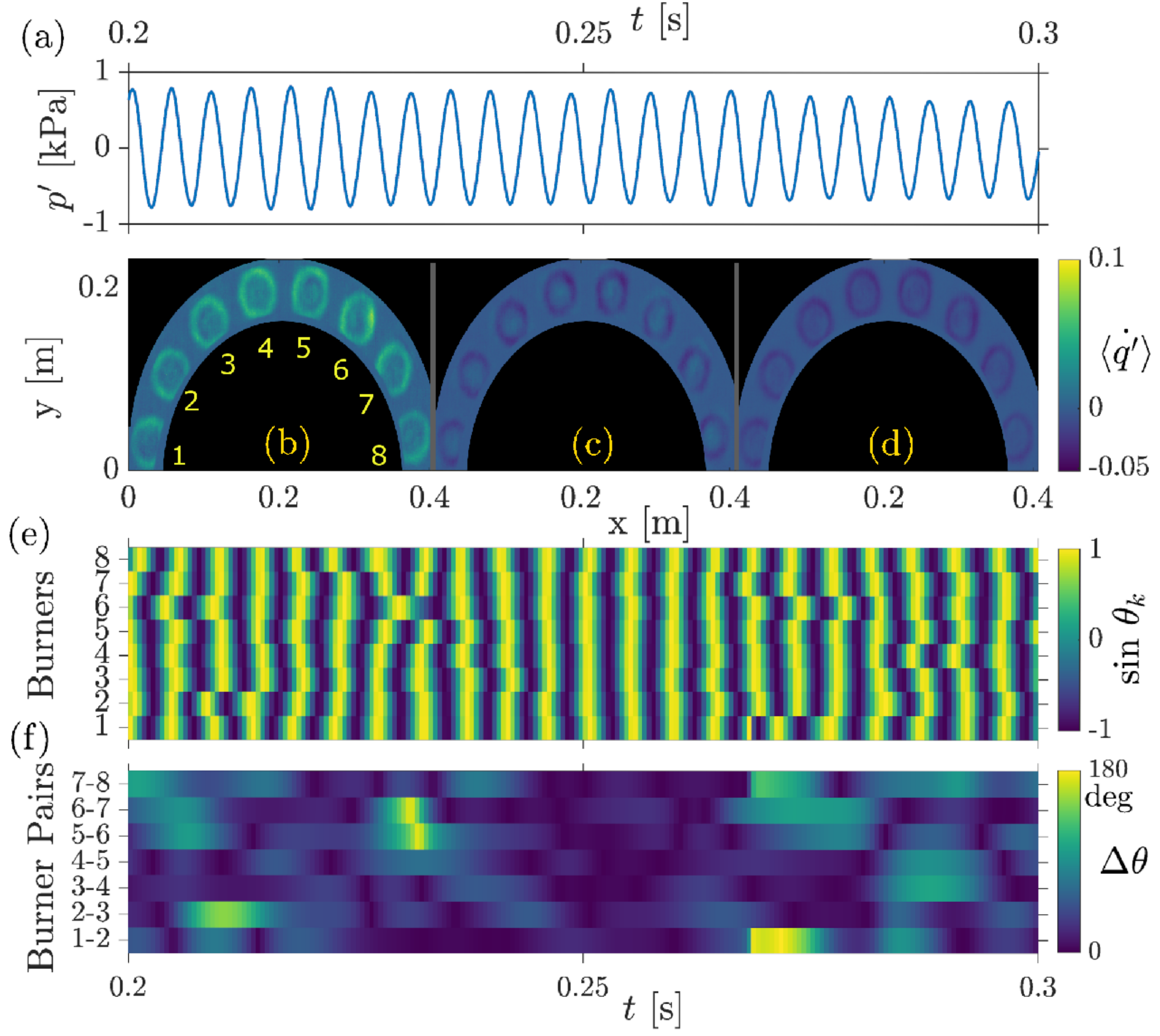}
\caption{\label{Fig5_low_amplitude_LCO} (a) Time series of $p^\prime$ during low amplitude TAI at $\phi=0.49$. Phase-averaged CH* images at pressure (b) maxima ($90^\circ$), (c) mean ($0^\circ$) and (d) minima ($270^\circ$) value. (e) Variation in the normalized amplitude of $\dot{q}^\prime$ for each burner. (f) Evolution of relative difference, $\Delta \theta$, between $\dot{q}^\prime$ from the indicated pairs of burners.}
\end{figure}

Figure \ref{Fig5_low_amplitude_LCO}a shows the time series of $p^\prime$ during low amplitude limit cycle with $p^\prime_{rms}\approx800$ Pa at $\phi=0.49$. The global flame structure can be observed from the phase-averaged CH* images obtained at the pressure maxima (90$^\circ$), mean ($0^\circ$), and minima ($270^\circ$) plotted in Figs. \ref{Fig5_low_amplitude_LCO}b-d, respectively. We can clearly observe a hollow flame structure for every burner along the annulus. The flame is bounded by the inner and outer shear layer with little to no recirculation. Consequently, there is a minima in the HRR at the centre of each flame and a large HRR along the flame edges. 

Next, we analyze the local flame behaviour and plot the normalized amplitude of HRR oscillations for each of the eight burners and phase difference between neighbouring burner in Figs. \ref{Fig5_low_amplitude_LCO}e \& f. We observe that the burners have the same frequency of HRR oscillations, as observed from the temporal match of their normalized amplitudes. We further find that the phase differences between neighbouring burners are predominantly close to zero, i.e., the burners are in-phase synchronized with each other. We also see phase slips appearing randomly between different pairs of burners (speck of bright spots). Phase slips indicate an increase in the phase difference between oscillators by $180^\circ$. We refer to this state where the flames are not perfectly synchronized as a state of weak synchronization.

\subsubsection{High amplitude thermoacoustic instability}
\begin{figure}[t!]
\centering
\includegraphics[width=0.75\textwidth]{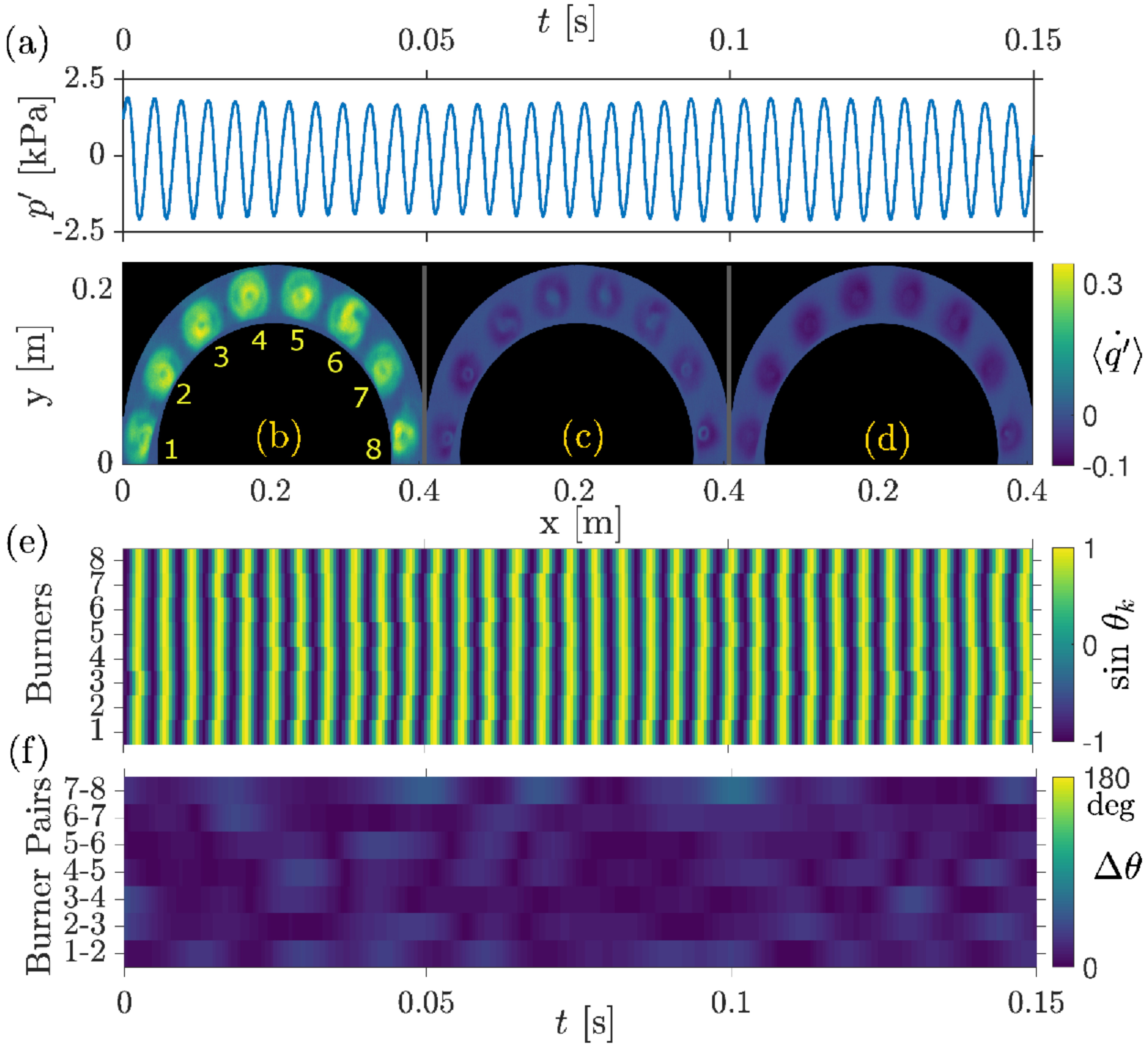}
\caption{\label{Fig6_high_amplitude_LCO} Flame dynamics observed during high amplitude TAI at $\phi=0.52$. Each subfigure is same as the last figure.}
\end{figure}

Figure \ref{Fig6_high_amplitude_LCO}a shows high amplitude TAI obtained at $\phi=0.52$. The amplitude of TAI is around 2 kPa and is about an order of magnitude larger than the low amplitude TAI. We plot the phase-averaged CH* images at the indicated phases in Figs. \ref{Fig6_high_amplitude_LCO}b-d, respectively. We observe that the flame dynamics are significantly different from that during low amplitude TAI. First, during pressure maxima, the highest HRR intensity is concentrated at the centre of each flame. {\color{black}This possibly indicates intense heat release as the flow recirculates into the inner recirculation zone during the pressure maxima. In contrast, at $0^\circ$ and $270^\circ$ phase, the flame does not propagate into the inner recirculation zone and remains confined to the shear layers.}

As before, we analyze the individual flames by evaluating the local HRR oscillations for each burner and compare the phase difference among them. In Fig. \ref{Fig6_high_amplitude_LCO}e, we observe that each of the burners attain maxima in the HRR at the same time instance, indicating in-phase synchronization among each of the burner pairs in addition to frequency synchronization. This is further corroborated from the temporal evolution of the phase difference between the pair of burners. We can observe that the burners are always in-phase synchronized, and the phase difference is always close to $0^\circ$. Hence, we refer to the flame interactions between burners during high amplitude TAI as perfect synchronization of all the burners.

\subsubsection{Quantitative analysis of synchronization characteristics}
Now, we quantify the relative degree of synchronization amongst different pair of burners and with the acoustic pressure oscillations. We define the phase-locking value (PLV) for any given pair of oscillators $x_1$ and $x_2$ as \cite{pikovsky2003synchronization, mondal2017synchronous}:
\begin{equation}
\textnormal{PLV} = \frac{1}{N}\bigg|\sum_{j=1}^N \exp\big(i\Delta \phi_{x_1,x_2}(t_j)\big)\bigg|,
\label{Eq-1}
\end{equation}
where, the phase difference between the signals at the instant $t_j$ is $\Delta \phi_{x_1,x_2}(t_j)=\phi_{x_1}(t_j)-\phi_{x_2}(t_j)$ and $N$ is the length of the time series. The PLV indicates the absolute value of the mean phase difference between two signals where the instantaneous phase differences ($\Delta\phi$) are expressed as complex unit-length vectors, i.e., $e^{i\Delta\phi}$ \citep{mondal2017synchronous}. The PLV has a value close to 0 for desynchronized signals and close to 1 for perfectly synchronized signals. For cases with partial synchronization such as intermittent phase-locking, the PLV is between 0 and 1. 

We also define the Kuramoto order parameter to quantify the synchronous behavior  for the spatially distributed oscillators (the eight burners) as \cite{mondal2017onset,dutta2019investigating}:
\begin{equation}
R(t) = \frac{1}{N_b}\bigg|\sum_{k=1}^{N_b}\exp(i\theta_k(t))\bigg|
\end{equation}
where, $\theta_k$ is the phase of the $k^{\text{th}}$ burner and $N_b$ is the total number of burners. At any time instance, $R=0$ indicates spatial desynchrony, while $R=1$ indicates spatial synchrony among the burners. 

\begin{figure}[t!]
\centering
\includegraphics[width=0.75\textwidth]{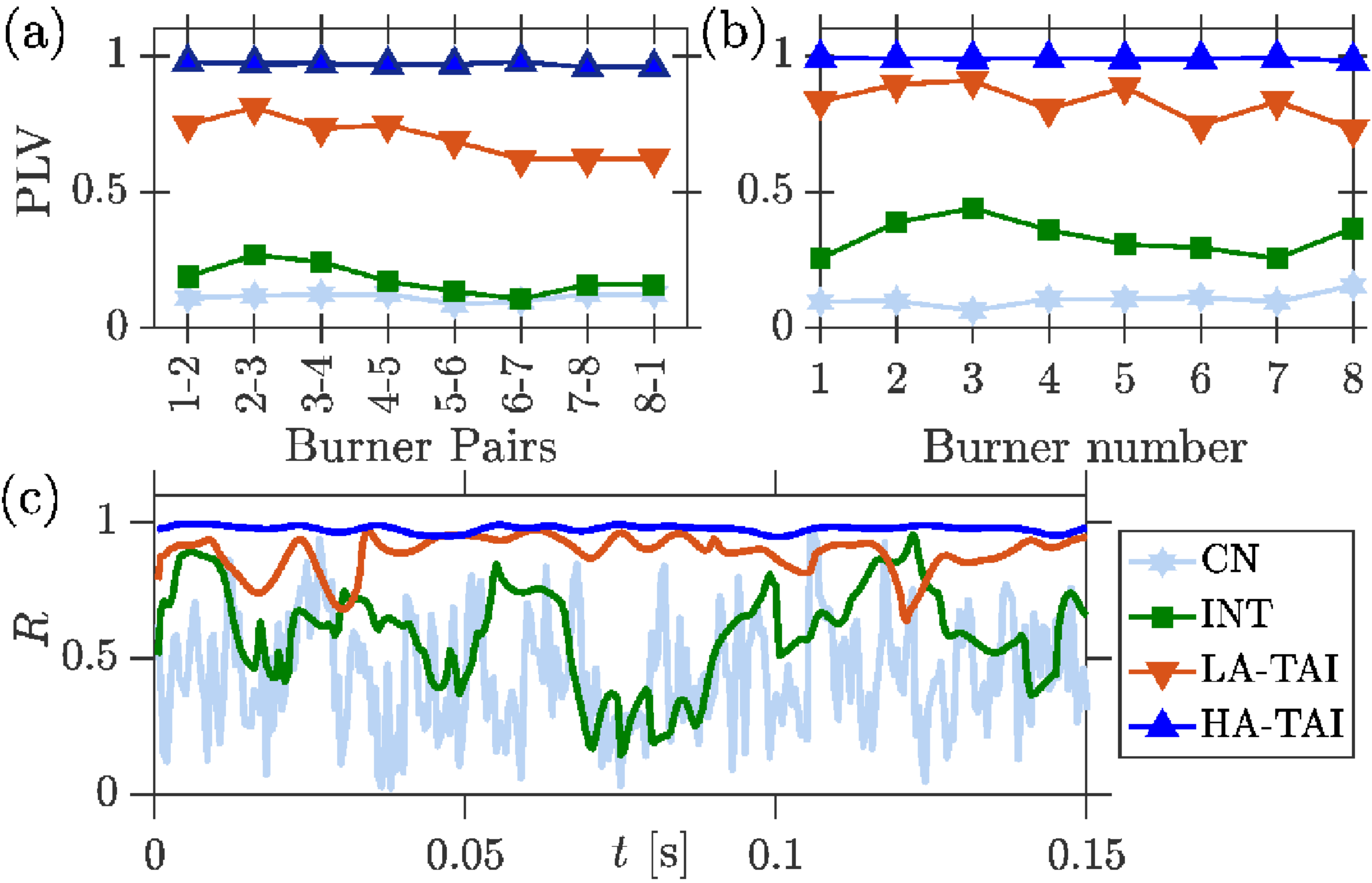}
\caption{\label{Fig7_PLV_Order_parameter} Phase-locking value (PLV) between (a) $\dot{q}^\prime$ measured from individual burners, and between (b) $\dot{q}^\prime$ from each burner and $p^\prime$ during combustion noise (CN), intermittency (INT) at $\phi=0.47$, low amplitude thermoacoustic instability (LA-TAI) at $\phi=0.49$, and high amplitude instability (HA-TAI) at $\phi=0.52$, respectively. (c) Kuramoto order parameter ($R$) determined from the eight burners during different states of combustor operation.}
\end{figure}

Figure \ref{Fig7_PLV_Order_parameter}a shows the variation of PLV between HRR oscillations from different pairs of neighbouring burners ($\dot{q}^\prime_{i,j}$) during the different dynamical states. Similarly, \ref{Fig7_PLV_Order_parameter}b shows the PLV between HRR oscillations of each burner ($\dot{q}^\prime_i$) with respect to $p^\prime$. During CN (Figs. \ref{Fig7_PLV_Order_parameter}a,b), the PLV among $\dot{q}^\prime_{i,j}$, and between $\dot{q}^\prime_i$ and $p^\prime$ remain close to zero, indicating desynchrony among burners and between burner and acoustics of the combustor. During INT, the PLV among $\dot{q}^\prime_{i,j}$ is very low ($<0.4$), indicating the desynchronized nature of their interaction with each other. However, the PLV between $\dot{q}^\prime_i$ and $p^\prime$ is close to 0.5, indicating partial synchronization between the burners and the acoustics.

During low amplitude TAI, flames are only weakly synchronized with each other due to phase-slips in their relative phases (Fig. \ref{Fig5_low_amplitude_LCO}f). As a consequence, PLV lies between 0.5 and 1, indicating weak synchronization among different burners (Fig. \ref{Fig7_PLV_Order_parameter}a). We also note that the PLV of different burners with $p^\prime$ follows suit and lies between 0.5 and 1 showing weak synchrony (Fig. \ref{Fig7_PLV_Order_parameter}b). For high amplitude TAI, the PLV among $\dot{q}^\prime_{i,j}$ and between $\dot{q}^\prime_i$ and $p^\prime$ lies close to 1, indicating perfect synchronization of the burners with each other and with the pressure oscillations (Figs. \ref{Fig7_PLV_Order_parameter}a,b). 

The Kuramoto order parameter is plotted as a function of time in Fig. \ref{Fig7_PLV_Order_parameter}c. The order parameter quantifies the temporal variation in the degree of spatial synchrony among different burners. During CN, $R$ fluctuates around time-averaged value of $\bar{R}=0.41$. During the periodic and aperiodic part of intermittency, $R$ fluctuates around $\bar{R}=0.49$ and $\bar{R}=0.37$. This indicates that spatially, the flames are partially synchronized during the periodic part of intermittency and desynchronized otherwise. The spatial desynchrony during aperiodic epochs of intermittency can be seen from the decrease below $R<0.5$. During low amplitude TAI, $R$ fluctuates around a mean value of $\bar{R}=0.84$. Thus, the burners are in a state of weak spatial synchronization. Finally, during high amplitude TAI, $R$ fluctuates around a mean value of $\bar{R}=0.97$, indicating perfect spatial synchronization.

{\color{black}The degree of spatio-temporal synchronization among burners during INT, low amplitude TAI and high amplitude TAI have important implications. During INT, the partial spatial synchronization leads to only partial fulfilment of the Rayleigh criteria as acoustic power sources where flame fluctuations are in-phase with $p^\prime$ cancel out due to acoustic sinks where flame fluctuations are out-of-phase with $p^\prime$. Thus, periodic oscillations are observed in intermittent bursts when a majority of the burners show spatial synchrony during the state of INT. During low amplitude TAI, burners are synchronized to a comparatively larger extent, and only random phase slips cause some burners to desynchronize (Fig. \ref{Fig5_low_amplitude_LCO}f). The phase slips indicate the momentary spatial desynchronization between certain burners and the combustor acoustics and affect the strength of acoustic power sources in the combustor. In contrast, during high amplitude TAI, the perfect synchronization between burners ensures that Rayleigh criteria are satisfied completely with acoustic power sources of significant strength distributed along the annulus and driving the high amplitude TAI. 

Finally, we note that the amplitude and phase response of the burners vary even when the flames are subjected to perturbations of similar amplitude during low amplitude longitudinal TAI. Further, the significantly different behaviour during high amplitude TAI is in keeping with the nonlinear dependence of flame response to dissimilar amplitude perturbation.}

\section{Conclusion}
{\color{black}In summary, we study the local and global flame dynamics observed during the transition from combustion noise (CN) to longitudinal thermoacoustic instability (TAI) in a sixteen burner swirl-stabilized lab-scale annular combustor. A systematic variation of equivalence ratio leads to the following states of combustor operation: CN, intermittency (INT), low amplitude TAI, and high amplitude TAI. We report the first observation of secondary bifurcation from low amplitude to high amplitude TAI in a turbulent annular burner. We contrast the flame structure observed during the various dynamical states. The flame structure changes from incoherent to well-defined ring-like structure with the flame stabilized along the shear layer during low amplitude TAI. Finally, during high amplitude TAI, intense heat release at the centre of each burner can be observed along the inner recirculation zone at the acoustic maxima.

Finally, we analyze the interactions between neighbouring flames along the annulus. Upon comparing amplitude and phase of HRR response of neighbouring burners, we find different degrees of spatio-temporal synchronization during different dynamical states.  We show that even for the case of longitudinal TAI, the flame-flame interactions are non-trivial. In particular, we find a transition from partially synchronized response of the burners during INT to weakly synchronized behavior with sporadic phase slips during low amplitude TAI, followed by perfect synchronization among the burners during high amplitude TAI. We quantify the degree of spatio-temporal synchronization using the phase-locking value and the Kuramoto order parameter. Most importantly, we characterize the nonlinear dependence of the flame response on the dissimilar amplitude perturbations encountered during low and high amplitude TAI.} 

\section*{Acknowledgments}
We thank Dr. Samadhan A. Pawar, Ms. Krishna Manoj, Ms. Reeja K. V., Mr. Midhun Raghunathan, Mr. S. Thilagaraj, and Mr. Anand for their help during this work. This work was supported by the Office of Naval Research Global (Contract Monitor: Dr R. Kolar) Grant no. N62909-18-1-2061.


\bibliographystyle{model1-num-names}
\bibliography{references}



\end{document}